\documentclass[prb,preprint,floatfix,amsmath,amssymb]{revtex4-1}

\usepackage{graphicx}
\usepackage{CJK}
\usepackage{comment}
\usepackage{lineno,hyperref}
\def\be{\begin{equation}}
\def\ee{\end{equation}}
\def\bea{\begin{eqnarray}}
\def\eea{\end{eqnarray}}
\usepackage{amsmath} 
\usepackage{bm}
\newcommand{\abs}[1]{\ensuremath{\lvert #1\rvert}}

\newcommand{\matrixel}[3]{\ensuremath{\left\langle #1 \vphantom{#2#3} \right| #2 \left| #3 \vphantom{#1#2} \right\rangle}}
\begin{document}

\title{The theory of transport in helical spin-structure crystals}
\author{Andrei Zadorozhnyi}
\affiliation{
Department of Physics and Astronomy/3905\\
1000 E. University Avenue\\ University of Wyoming\\ Laramie, WY 82071}
\author{Yuri Dahnovsky}
\affiliation{
Department of Physics and Astronomy/3905\\
1000 E. University Avenue\\ University of Wyoming\\ Laramie, WY 82071}
\email{yurid@uwyo.edu}

\date{\today}

\begin{abstract}
We study helical structures in spin-spiral single crystals. In the continuum approach for the helicity potential energy the simple electronic band splits into two non-parabolic bands. For $\varepsilon_{hel}$ greater than the splitting between the bands, the lower band is described by a surface with a saddle shape in the direction of the helicity axis. Using the Boltzmann equation with the relaxation due to acoustic phonons, we discover the dependence of the current on the angle between the electric field and helicity axis leading to the both parallel and perpendicular to the electric field components in the electroconductivity. In addition, we find that the transition rates depend on an electron spin allowing the transition between the bands. The electric conductivities exhibit nonlinear behaviors with respect to chemical potential $\mu$. We explain this effect as the interference of the band anisotropy, spin conservation, and interband transitions. The proposed theory with the spherical model in the effective mass approximation for conduction electrons can elucidate nonlinear dependencies that can be identified in experiments. There is the excellent agreement between the theoretical and experimental data for parallel resistivity depending on temperature at the phase transition from helical to ferromagnetic state in a MnP single crystal. In addition, we predict that the perpendicular resistivity abruptly drops to zero in the ferromagnetic phase.

\end{abstract}

\maketitle
\section{Introduction}

Helical spin structures take place in crystals because of relativistic effects that include spin-orbit coupling and anisotropy energy. \cite{ell,LL,calv1,robl,ign,xiao,ust} Mathematically equivalent problem occurs in neutron dynamics in helical magnetic field. \cite{calv2} Helical structures usually appear in phase diagrams along with other phases such as  ferromagnetic or antiferromagnetic states, skyrmions, etc. depending on temperature, magnetic field, and pressure. \cite{binz06} They take place, for example, in $\alpha-Cu V O_{7}$, \cite{gitg17} $MnSi$, \cite{jonietz2010} etc. Bcc $ Eu$ metals can also exhibit a spin-spiral state. \cite{olsen64} In the latter case the number of helical layers is  $n = 7$. To understand transport properties in spin-spiral antiferromagnetic crystals, we use the approach where free conduction electrons interact with the helical magnetic moments considered in the continuum approximation: \cite{calv1,robl, xiao}
\be\label{h0}
\hat{H}_0 = \hat{H}_{crys} + \hat{H}_{hel} = \frac{\hbar^2 k^2}{2m} - JS_{0}\left(\sigma_{x} \cos(\kappa z) + \sigma_{y} \sin(\kappa z)\right) = \frac{\hbar^2 k^2}{2m} - JS_{0}\hat{\bm{\sigma}}\cdot\bm{n}\left(\bm{r}\right),
\ee
where {\it J} is an exchange integral between the conduction electrons and localized magnetic moments, $S_0$. Here  $\sigma_{x,y}$ are the Pauli matrices. The helicity period  of the localized spin rotation about the z-axis is $2\pi/\kappa$. The discreet helicity potential in the tight binding approximation for the conduction electrons was considered in Ref. \cite{kipp21}. To explain experimental data, Hamiltonian (\ref{h0}) may not be sufficient. Fermi surfaces for realistic calculations can include non-parabolic $\varepsilon(\bm{k})$ and multi-band structure where the number of bands depends on the value of $\mu$. We use the simplified model to elucidate some effects that could be identified in experiments. The main purpose of this research is to study transport in helical system using the semiclassical approach based on the Boltzmann equation where a realistic scattering mechanism is taken into account. In particular, we consider electron-acoustic phonon scattering, which becomes spin-dependent in the helical potential energy. In addition, we prove that there is no a Berry curvature, and therefore, such materials are non-topological. We also prove that helical systems are insensitive to the chirality.  

\section{Berry curvature}
It is important to understand whether helimagnets are topological materials causing an abnormal Hall effect. To find a Berry curvature we use the Hamiltonian (\ref{h0}) that includes the helicity potential with the screw axis along $z$-direction. To diagonalize  the Hamiltonian, we use a unitary operator to satisfy the requirement 
\be
\hat{U}^{\dagger}\left( \hat{\bm{\sigma}}\cdot\bm{n}\right) \hat{U} = \hat{\bm{\sigma}} \cdot \bm{e}_{z}.
\ee
By direct substitution one can check that the following $\hat{U}$ fits:
\be 
\hat{U}  = \frac{1}{2}
\begin{pmatrix}
1  & e^{-i\varphi} \\
e^{i\varphi} & -1
\end{pmatrix},
\ee 
where $\varphi(z)$ is a polar angle for unit vector along the magnetization direction $\bm{n} =  \left(\cos{\varphi}, \sin{\varphi}, 0\right)$.

Using the unitary transformation  the Hamiltonian becomes: 

\be 
\begin{split}
	&\hat{U}^{\dagger}\hat{H}\hat{U} = \frac{\hat{U}^{\dagger}\bm{p}^2\hat{U}}{2m} - J\sigma_{z} =  \frac{\left(\hat{U}^{\dagger}\bm{p}\hat{U}\right)^{2}}{2m} - J\sigma_{z} \\
	&= \frac{\left(\bm{p} - i\hbar\hat{U}^{\dagger}\left(\bm{\nabla} \cdot \hat{U}\right)\right)^{2}}{2m} - J\sigma_{z} .
\end{split}
\ee 

Now we determine the gradient $\bm{\nabla} \hat{U}$:

\be 
\begin{split}
	&\bm{\nabla} \hat{U} = \frac{i}{2}\left(
	\begin{matrix}
		0 & - e^{-i\varphi}   \\
		 e^{i\varphi}& 0
	\end{matrix}
	\right) 	\bm{\nabla} \varphi.
\end{split}
\ee 

Then we find the product $\hat{U}^{\dagger} \bm{\nabla} \cdot \hat{U} = \hat{U} \bm{\nabla} \cdot \hat{U} $ as follows:

\be \label{unitary}
\begin{split}
	&\hat{U}^{\dagger} \bm{\nabla} \cdot \hat{U} 
	= \frac{i}{4} \left(
	\begin{matrix}
		1 & - e^{-i\varphi} \\
		- e^{i\varphi} & -1
	\end{matrix}
	\right) 	\bm{\nabla} \varphi
\end{split}
\ee 

In the adiabatic limit we neglect the off-diagonal matrix elements assuming that particle does not change the band.\cite{berry2011} Using this approximation and the first (\ref{unitary}) diagonal element we obtain the following  effective gauge field:

\be 
\bm{A}^{\uparrow\uparrow} = \bm{A}^{ad} = \frac{\hbar}{4e}(\bm{\nabla}\varphi). 
\ee 

As soon as the effective magnetic potential is a gradient, then the {\it curl} of it is zero. Thus the Berry curvature does not produce any effective magnetic field. 

\section{Electronic structure}

Hamiltonian $\hat{H}_0$ can be exactly diagonalized. This Hamiltonian was considered earlier in Refs. \cite{calv1,robl, xiao}. We provide the diagonalization procedure for the reader's convenience. 
To diagonalize the Hamiltonian we choose the $z$-axis along the helical one. The magnetic moments are in planes perpendicular to the helical axis. The directions of the magnetic moments are rotating about the helical axis with a period of $na=2\pi/\kappa$, where $a$ is the lattice constant. Schrodinger equation (\ref{h0}) can be solved by the separation of variables. Denoting $\Psi_x = e^{ik_x x}$, $\Psi_y = e^{ik_y y}$, and $\Psi_z^{\uparrow}$ and  $\Psi_z^{\downarrow}$ as unknown variables, where $\uparrow$ and $\downarrow$ signs denote the spin components, we find: 
\be \label{s_dif_eq}
\begin{split}
-\frac{\hbar^2}{2m}\left(\Psi_z^{\uparrow}\right)^{\prime\prime}_{zz}  - JS_{0} e^{-i\kappa z}\Psi_z^{\downarrow} = \varepsilon_{z}\Psi_z^{\uparrow},\\
-\frac{\hbar^2}{2m}\left(\Psi_z^{\downarrow}\right)^{\prime\prime}_{zz} - JS_{0} e^{+i\kappa z}\Psi_z^{\uparrow} = \varepsilon_{z}\Psi_z^{\downarrow},
\end{split}
\ee
where $\varepsilon_{z} = \varepsilon\left(\bm{k}\right) - \left(\hbar^2 k_x^2 + \hbar^2 k_y^2\right) / {2m}$. 
Then we present 

\be
\Psi_z^{\uparrow} = a e^{ik_z z} e^{-i\frac{\kappa}{2}z},\;\;\;\;\;\;\; \Psi_z^{\downarrow} = b e^{ik_z z} e^{i\frac{\kappa}{2}z},
\ee
where $\abs{a}^2 + \abs{b}^2 = 1$. Substituting these functions into Eq. (\ref{s_dif_eq}) we obtain:
\be \label{s_linear}
\begin{split}
a \frac{\hbar^2}{2m}\left( k_z - \frac{\kappa}{2} \right) ^2 - b JS_{0}  = a \varepsilon_{z},  \\
b \frac{\hbar^2}{2m}\left( k_z + \frac{\kappa}{2} \right) ^2 - a JS_{0}  = b \varepsilon_{z},
\end{split}
\ee
The solution of Eq. (\ref{s_linear}) is presented below:
\be \label{eps}
\varepsilon_{z}^{1, 2} =  \frac{\hbar^2 k_{z}^{2}}{2m}\pm \frac{\Delta }{2} \sqrt{ 1 +4 \frac{\varepsilon_{hel}}{\Delta^2} \frac{\hbar^2 k_{z}^{2}}{2m}  } + \frac{\varepsilon_{hel}}{4},
\ee 
where "-" corresponds to the lower band and "+" to the upper band. The splitting between bands at $k = 0$ is $\Delta = 2 J S_0$, and helicity energy is defined as $\varepsilon_{hel} ={\hbar^2 \kappa^{2}}/{2m}$. The coefficients $a$ and $b$ for eigenfunctions are:

\be \label{s_coef}
\begin{split}
a_{1} &= b_{2} =  \frac{1}{\sqrt{2}} \frac{\sqrt{ \frac{\Delta}{2} \sqrt{1 + 4\frac{\varepsilon_{hel}}{\Delta^2} \frac{\hbar^2 k_{z}^2}{2m}} + \frac{\hbar^2 }{2m}\kappa k_{z}  }}{ \sqrt{\frac{\Delta}{2}} \left( 1 +  4 \frac{\varepsilon_{hel}}{\Delta^2} \frac{\hbar^2 k_{z}^2}{2m} \right)^{1/4}},\\
a_{2} &= - b_{1} = - \frac{1}{\sqrt{2}} \frac{\sqrt{ \frac{\Delta}{2} \sqrt{1 + 4 \frac{\varepsilon_{hel}}{\Delta^2} \frac{\hbar^2 k_{z}^2}{2m}} - \frac{\hbar^2 }{2m}\kappa k_{z}  }}{ \sqrt{\frac{\Delta}{2}} \left( 1 +  4  \frac{\varepsilon_{hel}}{\Delta^2} \frac{\hbar^2 k_{z}^2}{2m} \right)^{1/4}}.
\end{split}
\ee

Then the eigenfunctions become:

\be \label{s_ef}
\begin{split}
\Psi^{1} = 
\left(
\begin{matrix}
	a_{1}\left(k_{z}\right) e^{-i\frac{\kappa z}{2}} \\
	b_{1}\left(k_{z}\right) e^{+i\frac{\kappa z}{2}}
\end{matrix}
\right) e^{ i \bm{k} \cdot \bm{r}}, \\
\Psi^{2} = 
\left(
\begin{matrix}
	a_{2}\left(k_{z}\right) e^{-i\frac{\kappa z}{2}} \\
	b_{2}\left(k_{z}\right) e^{+i\frac{\kappa z}{2}}
\end{matrix}
\right) e^{ i \bm{k} \cdot \bm{r}}.
\end{split}
\ee

\begin{figure}[htp]
\begin{center}
\includegraphics[width=8cm]{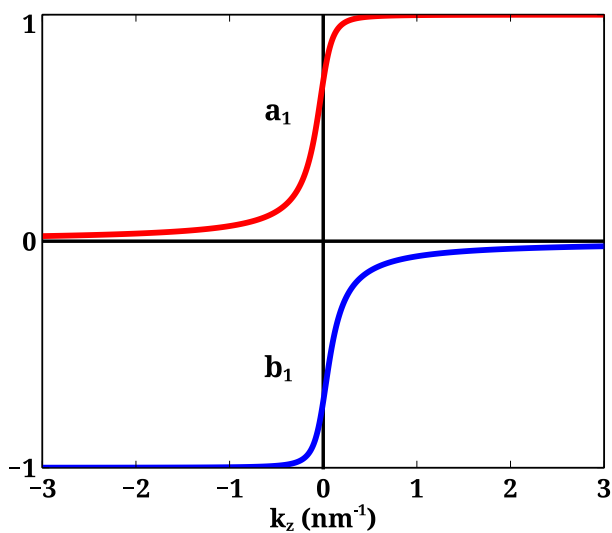}
\end{center}
\caption{\label{fig1}\small{The dependencies of coefficients, (a) spin-up $a_{1}$ and (b) spin-down $b_{1}$ of $k_{z}$ from Eq. (\ref{s_coef}) for the lower band.}} 
\end{figure}

The $k_z$-dependence of $a_1$ and $b_1$ is shown in Fig. \ref{fig1}. For $k_z > 0 $ $b_1\approx 0$ and $a_1\approx 1$. For $k_z < 0 $ $b_1\approx -1$ and $a_1\approx 0$. The intermediate region is about $\kappa \varepsilon_{hel} / \Delta$.

\begin{figure}[htp]
\begin{center}
\includegraphics[width=12cm]{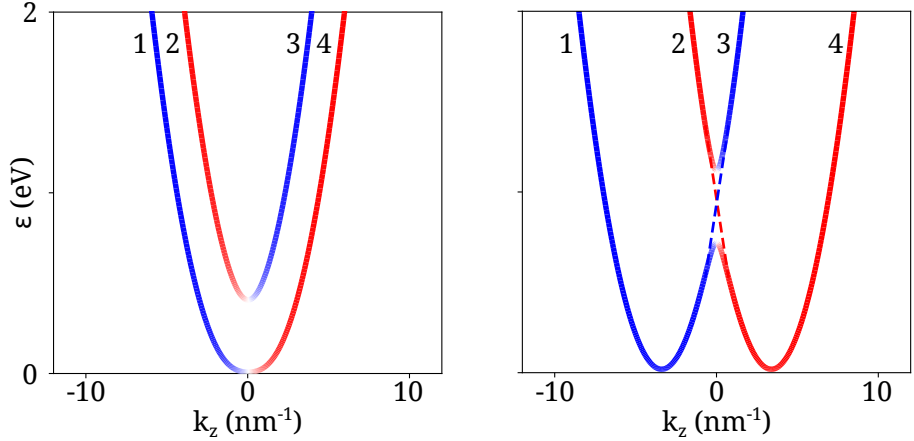}
\end{center}
\caption{\label{fig2}\small{Conduction electron energy bands in the $k_z$ direction for (a) $\varepsilon_{hel} < \Delta$ and (b) $\varepsilon_{hel} > \Delta$. The red and blue colors correspond to  the spin-up and spin-down states, respectively.}}. 
\end{figure}

From the mathematical analysis of the second derivatives of $\varepsilon_{1,2}$, we conclude that there are two possibilities for the shape of the lower band $\varepsilon_1$: (a) the single minimum curve for  $\varepsilon_{hel} < \Delta$ and  (b) the saddle shape  along the z-axis for $\varepsilon_{hel} > \Delta$ as demonstrated in Figs. \ref{fig2} {\it a} and {\it b}.  In the x,y-plane, the crossection  is still a paraboloid. The saddle shape is different compared to  Rashba interaction where $\varepsilon_1$ has a Mexican-hat form \cite{ZD2}. The upper band, $\varepsilon_2$, is always  of a single minimum shape. From the Hamiltonian diagonalization, we find the correspondence between the electron spin state and the $k_{z}$ component of the wave vector. In the lower band for $k_{z}>0$ the spin of the electron is $\uparrow$ (the red color), and for $k_{z}<0$ the spin is $\downarrow$ (the blue color). For the upper band, the dependence is the opposite. In the vicinity of $k_z \simeq 0$, there is a spin mixture. 

\section{Charge transport}

In this work we demonstrate that the transition rates are non-diagonal in 2D band space resulting in the transitions from one band to another. This effect is essential to explain the nonlinear behavior in current component perpendicular to the electric field. The main goal of this section is to determine transport properties in helical spin-structures. To do this, we use the semiclassical approach based of the nonequilibrium Boltzmann equation: \cite{anselm} 
\be\label{bol}
 \frac{\partial f_{0}}{\partial \varepsilon}e\bm{E}\cdot\bm{v}^{\nu}=\sum_{\nu ^{\prime}}\sum_{\bm{k}^{\prime}}\left(W_{\bm{k}\bm{k}^{\prime}}^{\nu \nu ^{\prime}}f_{1}^{\nu ^{\prime}}(\bm{k}^{\prime})-W_{\bm{k}^{\prime}\bm{k}}^{\nu ^{\prime} \nu}f_{1}^{\nu}(\bm{k})\right).
 \ee
The transition rates $W_{\bm{k}\bm{k}^{\prime}}^{\nu \nu ^{\prime}}=(2\pi/\hbar)\left|\matrixel{\bm{k}^{\prime} \nu ^{\prime}}{\hat{V}_{e-ph}}{\bm{k} \nu}\right|^{2} \delta(\varepsilon_{\nu}(\bm{k}) - \varepsilon_{\nu ^{\prime}}(\bm{k}^{\prime}))$, $f_0$ is the equilibrium Fermi distribution function, $f_{1}$ is the nonequilibrium part of the total distribution function, $\bm{E}$ is an applied electric field, and $\bm{v}$ is an electron velocity. Index $\nu $ denotes an energy band number ($\nu =1,2$). Bearing in mind that the system is a metal with no impurities of any kind, the electron scattering, in this case, results only from the electron-acoustic phonon interaction:
  \be \label{eph}
 \hat{V}_{e-ph} \approx - \bm{\nabla} \left(\hat{H}_{crys} + \hat{H}_{hel}\right) \cdot \bm{u}.
 \ee
 The same approach is valid for crystals with complex unit cell. Despite the presence of optical phonons, they do not contribute to the relaxation rates because the optical phonon frequency is higher than the temperatures considered in this work. For example, the estimated phonon energy in MnP is $h \nu = 0.0124\; \mathrm{eV}$ or $144 \; \mathrm{K} > T =  69 \; \mathrm{K}$. \cite{chong2016} 
$\hat{H}_{crys}$ and $\hat{H}_{hel}$ are defined in Eq. (\ref{h0}). The atom displacement, $\bm{u}$,  can be expressed in terms of the phonon normal coordinates. \cite{anselm} In Appendix we have presented the derivation of the 2D transition rate matrices with the  nonvanishing off-diagonal elements. $W_{\bm{k}\bm{k}^{\prime}}^{\nu\nu^{\prime}}$ can be analytically derived. 
 For calculations, we have used the Debye approximation where $\omega_{D} = v_{0}\left( 6 \pi^2 / \Omega_0 \right)$,  $v_0$ is a velocity of sound, and $\Omega_0$ is a unit cell volume. We have chosen a $Eu$ metal for numerical estimation where $v_s = 1860$ m/s and $\Omega_0 = 0.1 nm^{3}$. The sound velocity, $v_{s}$, has been estimated using the value of the Young modulus taken from  Ref. \cite{lide09}. According to Ref. \cite{anselm}, we employ the elastic scattering approximation for the numerical calculations where the transition rate matrix elements are given by the following equations:
 
 \be \label{w_fin}
\begin{split}
W_{\bm{k} \bm{k}^{\prime}}^{\nu\nu^\prime}&=\frac{2\pi}{\hbar} \left|\matrixel{\bm{k}^{\prime}, \nu^{\prime}, N_{\bm{q}j}^{\prime}}{\Delta V}{\bm{k}, \nu, N_{\bm{q}j}}\right|^{2} \delta(\varepsilon_{\nu}(\bm{k}) - \varepsilon_{\nu^{\prime}}(\bm{k}^{\prime})) \\
&= \frac{2\pi}{\hbar} \frac{1}{N M} \frac{ \hbar N_{\bm{q} j }}{ 2 \omega_{\bm{q} } } \left| K_{\nu \nu^{\prime}}^{+} \right|^{2} \delta(\varepsilon_{\nu}(\bm{k}) - \varepsilon_{\nu^{\prime}}(\bm{k}^{\prime})) \delta(\bm{k}^{\prime} - \bm{k} - \bm{q}) \\
&+ \frac{2\pi}{\hbar} \frac{1}{N M} \frac{ \hbar \left( N_{-\bm{q} j } + 1 \right)}{ 2 \omega_{\bm{-q} } } \left| K_{\nu \nu^{\prime}}^{-} \right|^{2} \delta(\varepsilon_{\nu}(\bm{k}) - \varepsilon_{\nu^{\prime}}(\bm{k}^{\prime}))\delta(\bm{k}^{\prime} - \bm{k} + \bm{q}),
\end{split}
\ee
where phonon wave vector $\bm{q} = \bm{k}^{\prime} - \bm{k}$ and $K_{\nu \nu^{\prime}}^{+}$ is determined by Eq. (\ref{kp_6}). $K_{\nu \nu^{\prime}}^{-}$ corresponds  to $\bm{k}^{\prime}=\bm{k}-\bm{q}$ and differs only in the sign in $K_{\nu \nu^{\prime}}^{+}$.
\be \label{kp_6}
\begin{split}
K_{\nu \nu^{\prime}}^{+} & = i \frac{\hbar ^ {2} }{2 m a^{2} } \frac{2}{3} \left( \bm{q} \cdot \bm{e}_{\bm{q} j } \right) \left( a^{\nu} \left(\bm{k}\right) a^{\nu ^{\prime}} \left(\bm{k}^{\prime}\right) +b^{\nu} \left(\bm{k}\right) b^{\nu ^{\prime}} \left(\bm{k}^{\prime}\right)  \right).
\end{split}
\ee

\begin{figure}[htp]
\begin{center}
\includegraphics[width=12cm]{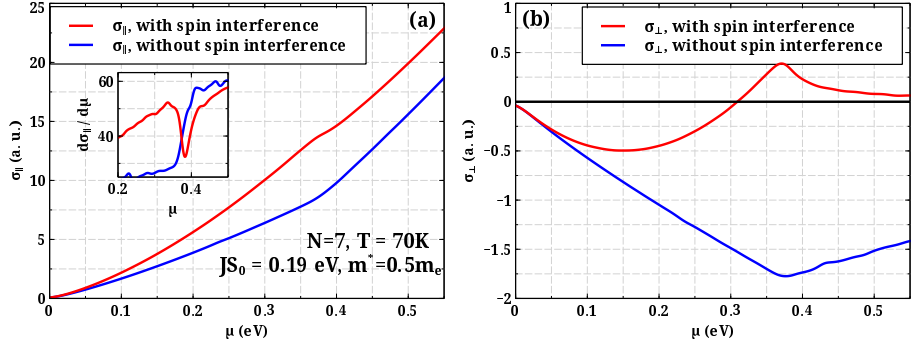}
\end{center}
\caption{\label{fig3}\small (a) Parallel and (b) perpendicular conductivities for  $\varepsilon_{hel} < \Delta$ (the single minimum lower band). The blue curves represent the conductivities for the spin-independent scattering and vanishing transitions between the bands and the red curves demonstrate the results of the full calculations. The insertion for the parallel component demonstrates $d\sigma_{\parallel}/d\mu$.} 
\end{figure}

Here $a^{\nu}$ and $b^{\nu}$ are determined by Eq. (\ref{s_coef}) and 
\be \label{s_bose}
\left<N_{\bm{q}j}\right> = \frac{1}{e^{{\varepsilon_{ph}}/{k_{B}T}} - 1}.
\ee
It is important to note at this point that for $\nu \ne \nu^{\prime}$ $K_{\nu \nu^{\prime}}^{+} \ne 0$. To solve the Boltzmann equation (\ref{bol}) with the electron-acoustic phonon scattering matrix, we have written the original codes where we have employed the nonuniform mesh. Then,  we  have numerically calculated the 3D integrals to determine the matrix elements. The main challenge is in computing the 3D integrals with the sharp $\delta$-function. As soon as $f_1$ has been found, we have determined the parallel and perpendicular conductivities assuming the angle between the $z$-axis and the electric field to be $\theta$. As shown in Ref. \cite{ZD1}, the parallel and perpendicular to the electric field components of electric current for an anisotropic crystal can be expressed as follows:
\be\label{j}
j^{s}_{\parallel} = j^{s}_{0 \parallel} + j^{s}_{2}\cos{\left( 2\theta\right) }, \;\;\;\;\;\;\;\;\;\;\;\;\;\;\;\;
j^{s}_{\perp} = j^{s}_{2}\sin{\left( 2\theta\right) }.
\ee
In these equations, the parallel component has the angle-independent part, $ j^{s}_{0 \parallel}$,  which, as follows from the calculations, is much greater than the amplitude $j^{s}_{2}$. Moreover, from the symmetry of the system in  the $x,y$-plane, the angle-independent part of the perpendicular component, similar to  $ j^{s}_{0 \parallel}$, vanishes. Because of the time-reversal symmetry, $\varepsilon(\bf{k})=\varepsilon(-\bf{k})$, the angular dependence is $2\theta$ instead of $\theta$. The results of the calculations  for electric conductivities are presented in Figs. \ref{fig3} and  \ref{fig5}. In this figures we study the conductivity versus chemical potential. The change in chemical potential can be implemented in experiments by the change of gate voltage applied in the $y$-direction. To avoid additional scattering mechanism due to charge impurities, we exclude doping as other way to change chemical potential. The lower energy band, $\varepsilon_1$, with the single minimum is depicted in Fig. \ref{fig2}a while the  case with the saddle shape for $\varepsilon_1$ is presented in Fig. \ref{fig2}b.   In Figs. \ref{fig3} and  \ref{fig5}, the parallel and perpendicular conductivities depend on chemical potential $\mu$ (see the red curves) for $\theta=\pi/4$.  For better understanding the nature of the transitions in the scattering rate matrix ${W}$ within the same energy band  and between the bands, we introduce the auxiliary model where we allow the transitions to take place only between the states within the same energy band. The spin-dependence is not considered either.  Such a procedure allows us to exclude the interference due to the spin dependence within a single band. As shown in Fig. \ref{fig2}, the transitions within one band from $k_z>0$ to $k_z<0$ are practically forbidden except the states lying in the vicinity   of $k_z=0$. Indeed, $\left( a^{\nu} \left(\bm{k}\right) a^{\nu ^{\prime}} \left(\bm{k}^{\prime}\right) +b^{\nu} \left(\bm{k}\right) b^{\nu ^{\prime}} \left(\bm{k}^{\prime}\right)  \right)$ in Eq. (\ref{kp_6}) is a scalar product of the spin states with $\bm{k}$ and $\bm{k}^{\prime}$. For larger $k_{z}$, these states are either $(1, 0)$ ($k_{z}>0$) or $(0, 1)$  ($k_{z}<0$). Then, the product, $ a^{\nu} \left(\bm{k}\right) a^{\nu ^{\prime}} \left(\bm{k}^{\prime}\right) +b^{\nu} \left(\bm{k}\right) b^{\nu ^{\prime}} \left(\bm{k}^{\prime}\right) =0$, resulting in $W_{\bm{k} \bm{k}^{\prime}}^{\nu\nu^\prime}=0$. Thus, the active phase volume for the electron scattering is reduced approximately by a factor of $2$, resulting in greater conductivity, as shown in Fig. \ref{fig3}a. The insertion in Fig. \ref{fig3}a demonstrates the sharp minimum in $d\sigma_{\parallel}/d\mu$. 

\begin{figure}[htp]
\begin{center}
\includegraphics[width=10cm]{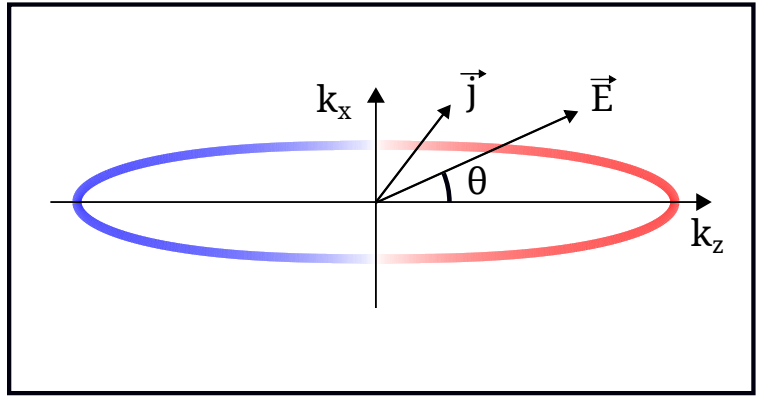}
\end{center}
\caption{\label{fig4}\small{Energy surface cross-section in $k_x$, $k_z$ plane for the lower band. $\theta$ is an angle between the applied electric field and the $z$-axis. The red and blue colors correspond to spin-up and spin-down states, respectively.}}
\end{figure} 

From the analysis of Eq. (\ref{eps}), we find that the effective masses for both bands become anisotropic. Indeed, $m_z > m$ corresponds to the lower band and  $m_z < m$ is for the upper band. The anisotropy of the bands is schematically depicted in Fig. \ref{fig4}. In the case of the band anisotropy, it was proven in Ref. \cite{ZD1} that the current has a perpendicular component if the angle between $z$-axis and an electric field is nonzero (see Eq. (\ref{j})). 

For the perpendicular component of the electric conductivity (see Fig. \ref{fig3}b), we find the similar dependencies. Indeed, the current initially grows in amplitude. However, the direction of the current is negative, i. e., leaning towards the $x$-axis because $m_{x} < m_{z}$. The similar behavior is observed for $\sigma_{\perp}$ in the auxiliary model for low energies. Such a dependence in the vicinity of $k_z = 0$ can be explained by the transitions between the electronic states where the wavefunctions are the linear combinations of the spin-up and spin-down states. In this case, the active region of the $k$-space is the same for both models. Then, the dependencies strongly diverge. For the general model (the red curves in Fig. \ref{fig3}), the amplitude of  $\sigma_{\perp}$ decreases while for the auxiliary model the trend toward the greater negative values continues (the blue curve). It happens because in the realistic model, the half of the phase volume becomes unavailable for the electron scattering. When $\mu$ reaches the bottom of the upper band (it is about $2.2\;eV$ in Fig. \ref{fig3}a), there is a drop in $\sigma_{\perp}$ and the slower growth in $\sigma_{\parallel}$ (see the insertion in Fig. \ref{fig3}a for the derivative). We explain the slow down in growth by the interband transition of the carriers with high velocities  from the lower band ($\varepsilon_1$) to the electronic states of the upper band ($\varepsilon_2$) where the electron velocities are small, and therefore, slightly contributing  to the total current. It is important to note that the peak or plateau in $\sigma_{\perp}$  and  $\sigma_{\parallel}$ (the red lines) are the result of the nondiagonal transitions in the scattering rate matrix $W$. Indeed, when these transitions are omitted (the  blue curves), we do not find any peaks. The perpendicular component changes the sign leaning towards the $z$-axis with the inclusion of the upper band carriers ($m_z<m_x$). 

\begin{figure}[htp]
\begin{center}
\includegraphics[width=12cm]{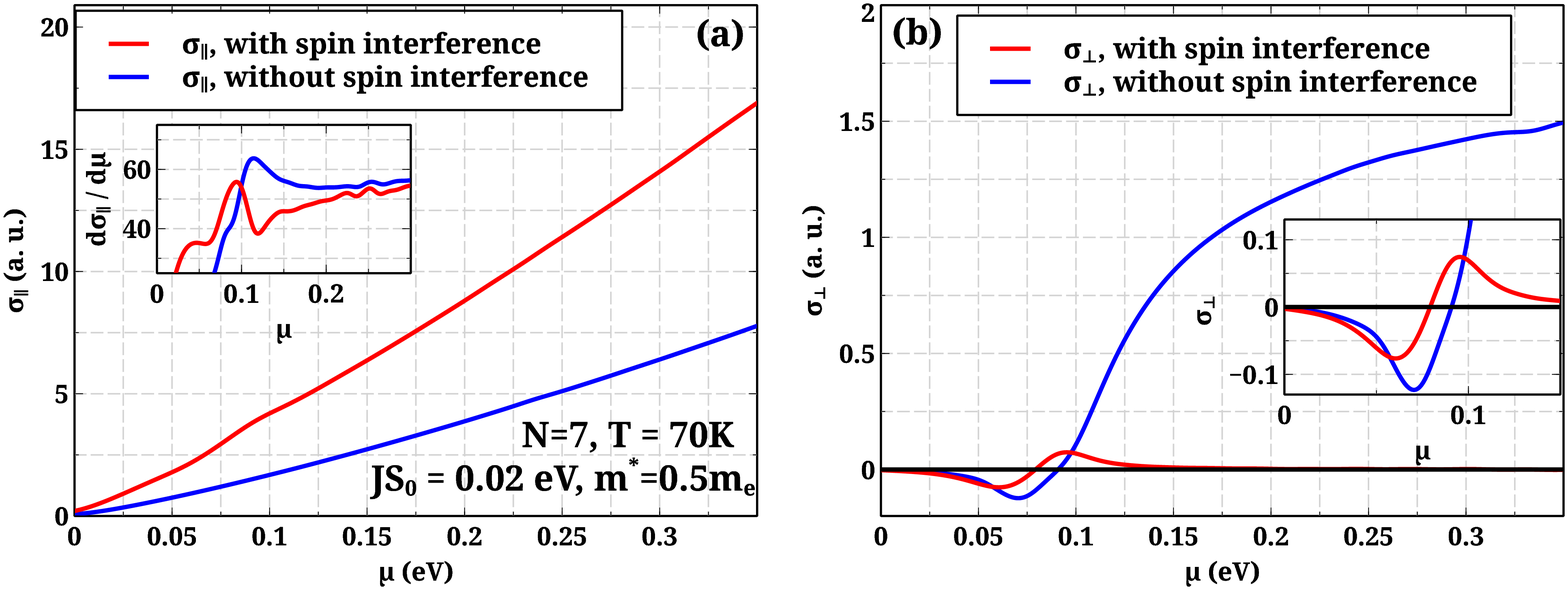}
\end{center}
\caption{\label{fig5}\small{(a) Parallel and (b) perpendicular conductivities  for $\varepsilon_{hel} > \Delta$ (the saddle shape case). The blue curves represent the conductivities for the spin-independent scattering and vanishing transitions between the bands and the red curves demonstrate the results of the full calculations. The insertion for the parallel component shows $d\sigma_{\parallel}/d\mu$.}}
\end{figure} 
The perpendicular component $\sigma_{\perp}$ behaves in the similar way as shown in Fig. \ref{fig3}b. The minimum in $\sigma_{\perp}$ (Fig. \ref{fig5}b, the red curve) can be explained by the saddle point in $\varepsilon_{1}(\bm{k})$. In this case the transitions within $\varepsilon_{1}$ take place from higher $k_{z}$ to the lower ones close to the saddle point. When the electron states are close to the saddle point, the velocities of electrons are very small and therefore, slightly contribute to the electric current. This is the reason why we find the drop of the absolute value of $\sigma_{\perp}$. The same behavior is observed for the auxiliary model. The existence of the maximum is the result of the switching on the upper band $\varepsilon_2$. Then, we find a very unusual dependence where $\sigma_\perp$, described by the red curve, is very small in amplitude compared to the blue curve in the auxiliary model. Indeed, at large $k_{z}$s ($\varepsilon_{hel} \hbar^2 k_{z}^2  / 2 m \gg \Delta^2/4$) it is possible to approximate 
\be\label{na}
\varepsilon_{1,2}^{na} \approx \frac{\hbar^2}{2 m} \left( k_{x}^2  + k_{y}^2 +  \left( k_{z} \pm \kappa/2\right)^2 \right) ,
\ee
 resulting in the nonadiabatic basis set. These parabolas are depicted in Fig. \ref{fig2}b as (1, 3) and (2, 4), where the spin on each parabola is conserved. In the nonadiabatic basis set the transitions take place within each parabola. For each nonadiabatic curve in Eq. (\ref{na}) there is no anisotropy. In this case $j_2^{s}$ in Eq. (\ref{j}) is close to zero explaining the red curve behavior at large $\mu$. For the auxiliary model, the basis set is still adiabatic, i. e., only transitions (1-4) and (2-3) take place. The original bands ((1-4) and (2-3)) are highly anisotropic and therefore, the perpendicular component is much greater.

We have found no chirality in helical systems because the transition rates are the even functions with respect to $\kappa$, i. e., $\bf{j}(+\kappa)=\bf{j}(-\kappa)$. This general statement has also been numerically verified.

\section{Temperature dependence}

The proposed theory allows to study the temperature dependence of conductivity. For the calculations, we have chosen the temperature to be $T\approx 69$ K. We do not consider very low temperatures because in a realistic situation  other scattering mechanisms such as scattering by impurities, will prevail over the electron-acoustic phonon scattering. At higher temperatures, there might be a phase transition into another phase state rather than a helical one. The temperature comes into the electron Fermi distribution function in the expression for the electric current. Besides the Fermi distribution function,  there is the phonon  Bose-Einstein distribution that appears in the transition rate expression due to the acoustic phonons (see Appendix).  At $T = 69$ K (the high temperature limit for bosons), $W \sim T$, consequently, $\sigma \sim 1/T$ or $\rho_{\parallel}\sim T$  \cite{anselm}
\begin{figure}[htp]
	\begin{center}
		\includegraphics[width=10cm]{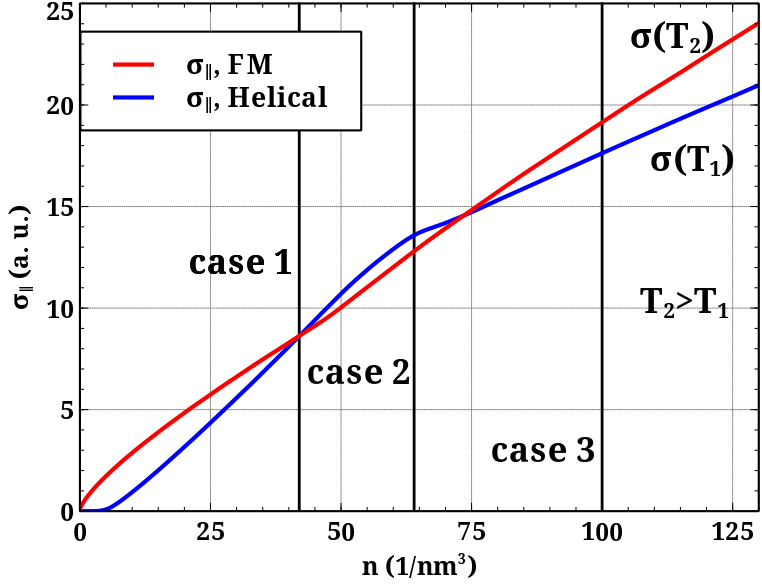}
	\end{center}
	\caption{\label{fig6}\small{Parallel conductivity for $\varepsilon_{hel} < \Delta$ with respect to the free electron density. The blue curve represents helical magnetic structure and the red curve represents ferromagnetic material. $JS_0 = 0.2$ eV, $m^* = 0.5 m_e$, $N = 7$, $k_{B}T = 0.006$ eV.}}
\end{figure} 

There are some experimental data where it is possible to check the proposed theory at structural phase transitions from  helical to ferromagnetic state. Indeed, such transitions take place in (a) MnP, $T_{c} = 50$ K\cite{zheng2017, jiang2020}, (b) $\mathrm{YMn_6Sn_6}$, $T_{c} = 333$ K\cite{ghimire2020}, and (c) MnCoSi, $T_{c}=110$ K \cite{zhang2008}. In case (a) there is the observable discontinuity in the conductivity, while in cases (b) and (c) there are the slope changes at the transition temperatures. The theory proposed above is able to explain such nonanalytic behaviors in $\sigma(T)$. Such effects are demonstrated in Fig. \ref{fig6}. In this graph, $\sigma$ is shown with respect to the number of electrons rather than chemical potential $\mu$. It is important to note that at phase transitions the chemical potential can have discontinuity due to the change of effective mass, and therefore, it is not a proper parameter to describe the conductivity dependencies. The quantity that is conserved at phase transition is  electron concentration. As shown if Fig. \ref{fig6}, there are three possible cases where at the fixed number of electrons the conductivity change takes place as a discontinuity with a positive (case 3) or negative (case 2) value or remains the same (case 1). In the latter case we assume that the temperature dependencies will provide the slope change in $\sigma(T)$. 

\begin{figure}[htp]
	\begin{center}
		\includegraphics[width=12cm]{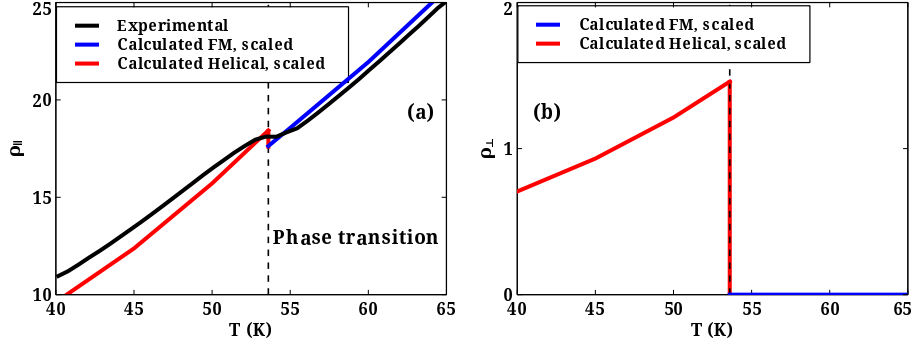}
	\end{center}
	\caption{\label{fig7}\small{Temperature dependence of resistivity, (a)  parallel and (b) perpendicular to the electric field components. The red and blue curves are theoretical and the black curve in (a) is experimental (see Ref. \cite{jiang2020}). $JS_{0} = 0.02$ eV, $m^{*} = 0.5 m_{e}$, $n = 2 \times 10^{18}$ cm$^{-3}$, the period of the spiral is $9$ lattice constants. }}
\end{figure}

In Fig. \ref{fig7}a and b, we demonstrate the temperature dependence of the parallel and perpendicular components of the resistivity at the phase transition from helical to ferromagnetic states for MnP, respectively ($T_c\approx50$ K).\cite{jiang2020} For the calculations we have chosen the following values of the parameters: $JS_{0} = 0.02$ eV, $m^{*} = 0.5 m_{e}$, $n = 2 \times 10^{18}$ cm$^{-3}$, the period of the spiral is $9$ lattice constants. 

As shown in Fig. \ref{fig7}a, the experimental (the black curve) and calculated (the red and blue curves) temperature dependencies of the resistivity are in very good agreement. At $T_{c} \approx 50$ K we observe a discontinuity in the calculated dependencies and smoothened discontinuity in experimental (black) curve. Along with the explanation of the experimental data, we have predicted the temperature dependence of the perpendicular component for the same values of the parameters (see Fig. \ref{fig7}b). In the helical phase we observe the growing resistivity, which abruptly vanishes at the phase transition. It is oblivious that the perpendicular resistivity is absent in the ferromagnetic state because of the spherical symmetry of $\varepsilon \left(\bm{k}\right)$. 

\section{Conclusions}

In this research we propose the theory to study transport properties of helimagnets. The system is described by Hamiltonian (\ref{h0}) with free electron potential for conduction electrons. Such approximation may not be sufficient for the complete explanation of experimental dependencies. However, it does elucidate the effects predicted in this research, i. e., the existence of the perpendicular to the field electric conductivity where maxima and minima with respect to $\mu$ can be found. The general theory should include realistic Fermi surfaces, non-parabolic $\varepsilon(\bm{k})$ and multi-band structure where the number of bands depends on $\mu$. We have discovered that the Berry curvature vanishes, and therefore, there is no abnormal spin-Hall effect. As a result of the helicity potential, the band splits into the two bands  \cite{calv1,robl, xiao} described by Eq. (\ref{eps}) where the lower band can be a single-minimum curve for $\varepsilon_{hel} < \Delta$ or a saddle shape for $\varepsilon_{hel}>\Delta$ (see Fig. \ref{fig2}). These bands are anisotropic resulting in the origin of the parallel and perpendicular components of the electric currents according to Eq. (\ref{j}).   Electron-acoustic phonon interaction has been chosen as a scattering mechanism. We have proven that the scattering between the electronic bands is allowed and the transition rates are spin-dependent. Both components reveal nonlinear behaviors with respect to $\mu$ as demonstrated in Figs. \ref{fig3} and \ref{fig5}. The change in chemical potential can be implemented in experiments by the change of gate voltage applied in the $y$-direction. To avoid additional scattering mechanism due to charge impurities, we exclude doping as other way to change chemical potential. The perpendicular to the electric field electroconductivity exhibits the unusual dependencies. Such behaviors have been explained in terms of the anisotropy mass model combined with the $k$-space restricted by the spin conservation. The proposed theory is capable of the explanation of the nonanalytic behaviors in some materials where discontinuities and slope changes in conductivities take place at phase transitions from helical to ferromagnetic states. \cite{zheng2017, jiang2020, ghimire2020, zhang2008} Indeed, as shown in Fig. \ref{fig7}, we have found the excellent agreement between the experimental and calculated dependencies of the parallel to the electric field resistivity at the phase transition for MnP at $T_{c} \approx 50$ K. In addition, we have predicted the abrupt behavior of the perpendicular resistivity, which drops to zero in the ferromagnetic phase.

\section*{Acknowledgment}
This work was supported by a grant from the U S National Science Foundation (No. DMR-1710512) and the U S Department of Energy (No. 1004389) to the University of Wyoming.

\section*{Appendix}
\appendix
\renewcommand{\theequation}{A.\arabic{equation}}

To solve the Boltzmann equation, we need to find transition rates $W$, which are determined as follows: \cite{anselm}\be \label{s_w}
W_{\bm{k} \bm{k}^{\prime}}^{\nu\nu^\prime}=(2\pi/\hbar)\left|\matrixel{\bm{k}^{\prime}, \nu^{\prime}, N_{\bm{q}j}^{\prime}}{\Delta V}{\bm{k}, \nu, N_{\bm{q}j}}\right|^{2} \delta(\varepsilon_{\nu}(\bm{k}) - \varepsilon_{\nu^{\prime}}(\bm{k}^{\prime})),
\ee
where $\nu$ and $\nu^{\prime}$ denote band indecis ($\nu = 1, 2$).The electron scattering is determined by the interaction of electrons with acoustic phonons.

For the matrix element 

\be \label{s_mat_el}
\matrixel{\bm{k}^{\prime}, \nu^{\prime}, N_{\bm{q}j}^{\prime}}{\Delta V}{\bm{k}, \nu, N_{\bm{q}j}},
\ee
we denote $\bm{k}$ as an electron wave vector, and $ N_{\bm{q}j} $ is phonon population number with the wavevector $\bm{q}$ and the branch $j$. For electron wavefunctions, we select the adiabatic basis set determined by Eq. (\ref{s_ef}). Expression (\ref{s_mat_el}) allows the transitions between the bands ($\nu \ne \nu^\prime$). 

The linear expansion with respect to the lattice vibrations of the total potential energy that includes the crystal and helicity terms yields:

\be
\begin{split}
& \Delta V_{s s^{\prime}} = V_{s s^{\prime}}\left(\bm{r}\right) - V_{s s^{\prime}}\left(\bm{r} + \bm{u}\right) = - \bm{u} \bm{\nabla} V_{s s^{\prime}}, \\
& \bm{u} = \frac{1}{\sqrt{NM}} {\sum_{qj}}^{\prime} \bm{e}_{\bm{q}j} \left(a_{\bm{q}j} e^{i \bm{q} \bm{r}} + a_{\bm{q}j} ^{*} e^{- i \bm{q} \bm{r}}\right), 
\end{split}
\ee
where $s, s^{\prime}$ are the spin projection indices. The displacement $\bm{u}$ is presented in terms of the normal coordinates. For acoustic phonons, $M$ is a mass of the unit cell, "$\prime$" means that summation takes place only over the $half$ of the Brillouin zone, $V$ is the potential of the unperturbed lattice, and $\Delta V$ is perturbed potential in linear expansion. \cite{anselm} The matrix element (\ref{s_mat_el}) can be presented in terms of the phonon normal coordinates in the following way: 
\be \label{s_mat_el_explicit}
\begin{split}
	&\matrixel{\bm{k}^{\prime}, \nu^{\prime}, N_{\bm{q}j}^{\prime}}{\Delta V}{\bm{k}, \nu, N_{\bm{q}j}} \\
	&= - \frac{1}{\sqrt{N M }} \sum_{s, s^{\prime}}  \int \left[\psi^{\nu^{\prime} *}_{\bm{k}^{\prime} s^{\prime}} \left(\bm{r}\right) \prod_{\bm{q} j} \Psi^{*}_{N^{\prime}_{\bm{q}j}} \left( Q_{\bm{q} j} \right) (\bm{\nabla} V) \cdot {\sum_{\bm{q}j}}^{\prime} \bm{e}_{\bm{q}j} \left(a_{\bm{q}j} e^{i \bm{q} \bm{r}} + a_{\bm{q}j} ^{*} e^{- i \bm{q} \bm{r}}\right)  \right.
	\\
	&\left. \times  \psi^{\nu}_{\bm{k}s} \left(\bm{r}\right) \prod_{\bm{q} j} \Psi_{N_{\bm{q}j}} \left( Q_{\bm{q} j} \right) \right] d\tau \prod_{\bm{q} j}  dQ_{\bm{q} j}.
\end{split}
\ee
In this expression, $d\tau$ is the infenitesimal  volume in the electron coordinates, $Q_{\bm{q}j}$ are normal phonon coordinates, $\bm{e}_{\bm{q}j}$ is a phonon polarization vector, and $\psi^{\nu}_{\bm{k} s} \left(\bm{r}\right) = u^{\nu}_{\bm{k} s}  \left(\bm{r}\right) e^{i \bm{k} \cdot \bm{r}}$ are electron and $\Psi_{N_{\bm{q}j}} \left( Q_{\bm{q} j} \right) $ are phonon wave functions, respectively. The integration in Eq. (\ref{s_mat_el_explicit}) can be performed for the electron coordinates and the phonon coordinates separately:
\be \label{s_mat_el_expanded}
\begin{split}
	&\matrixel{\bm{k}^{\prime}, \nu^{\prime}, N_{\bm{q}i}^{\prime}}{\Delta V}{\bm{k}, \nu, N_{\bm{q}i}} \\
	= - \frac{1}{\sqrt{N M }}\sum_{ss^{\prime}} {\sum_{\bm{q}j}}^{\prime} \Bigg\{ &\left[ \frac{1}{N} \bm{e}_{\bm{q} j } \cdot \int (\bm{\nabla} V) e^{i \left(\bm{k} + \bm{q} - \bm{k}^{\prime}\right) \cdot \bm{r}} u^{\nu^{\prime}*}_{\bm{k}^{\prime} s^{\prime}}  \left(\bm{r}\right) u^{\nu}_{\bm{k} s}  \left(\bm{r}\right) d\tau\right]\Bigg. 
	\\
	\times&\left[  \int \prod_{\bm{q} j} \Psi^{*}_{N^{\prime}_{\bm{q}j}} \left( Q_{\bm{q} j} \right)  a_{\bm{q}j}  \prod_{\bm{q} j} \Psi_{N_{\bm{q}j}} \left( Q_{\bm{q} j} \right) \prod_{\bm{q} j}  dQ_{\bm{q} j}\right]
	\\
	  +  &\left[ \frac{1}{N} \bm{e}_{\bm{q} j } \cdot \int (\bm{\nabla} V) e^{i \left(\bm{k} - \bm{q} - \bm{k}^{\prime}\right) \cdot \bm{r}} u^{\nu^{\prime}*}_{\bm{k}^{\prime} s^{\prime}}  \left(\bm{r}\right) u^{\nu}_{\bm{k} s}  \left(\bm{r}\right) d\tau \right]
	\\
	\times &\Bigg. \left[ \int \prod_{\bm{q} j} \Psi^{*}_{N^{\prime}_{\bm{q}j}} \left( Q_{\bm{q} j} \right)  a^{*}_{\bm{q}j}  \prod_{\bm{q} j} \Psi_{N_{\bm{q}j}} \left( Q_{\bm{q} j} \right) \prod_{\bm{q} j}  dQ_{\bm{q} j} \right] \Bigg\},
\end{split}
\ee
 The expressions in the first, second, third, and forth brackets are denotes as  $K_{\nu \nu^{\prime}}^+$, $L$, $K_{\nu \nu^{\prime}}^{-}$, and $L^+$, respectively. For $L$, using quantum oscillator properties for lowering and raising operator, we find non-vanishing matrix elements for $L$ and $L^+$:
\be
\begin{split}
L &= \matrixel{N_{\bm{q}j} - 1}{a_{\bm{q}j}}{N_{\bm{q}j} } = \sqrt{\frac{\hbar N_{\bm{q}j} }{2 \omega_{\bm{q} j}}},\\
L ^+ &= \matrixel{N_{\bm{q}j} + 1}{a_{\bm{q}j}}{N_{\bm{q}j} } = \sqrt{\frac{\hbar \left( N_{\bm{q}j} + 1 \right) }{2 \omega_{\bm{q} j}}}.
\end{split}
\ee 
For the electron coordinate integration:
\be \label{s_kp_1}
K_{\nu \nu^{\prime}}^{+} = \frac{1}{N} \sum _{s, s^{\prime}} \sum_{n} e^{i \left( \bm{k} + \bm{q} - \bm{k}^{\prime} \right) \bm{a}_{n}} \int e^{i\left( \bm{k} + \bm{q} - \bm{k}^{\prime} \right) \bm{r}} \left( \bm{e}_{ \bm{q} j } \cdot \bm{\nabla} V _{s^{\prime} s} \left(\bm{r}\right) \right) u_{ s \bm{k} }  ^ { \nu }  \left( \bm{r} \right) u_{ s^{\prime} \bm{k}^{\prime} } ^ { \nu^{\prime} \ast} \left( \bm{r} \right) d\tau_{0}
\ee
From the detailed analysis it can be proven that $V _{s^{\prime} s} \left(\bm{r}\right) u_{ s \bm{k} }  ^ { \nu }  \left( \bm{r} \right) u_{ s^{\prime} \bm{k}^{\prime} } ^ { \nu^{\prime} \ast} \left( \bm{r} \right)$ is a periodic function with respect to the unit lattice vector. 

The summation over $n$ in Eq. (\ref{s_kp_1}) yields the momentum conservation: $\bm{k}^{\prime} = \bm{k} + \bm{q}$, resulting in 
\be \label{s_kp_2}
\begin{split}
	K_{\nu \nu^{\prime}}^{+} = \sum _{s, s^{\prime}} \int \left( \bm{e}_{ \bm{q} j } \cdot \bm{\nabla} V _{s^{\prime} s} \left(\bm{r}\right) \right) u_{ s \bm{k} }  ^ { \nu }  \left( \bm{r} \right) u_{ s^{\prime} \bm{k}^{\prime} } ^ { \nu^{\prime} \ast} \left( \bm{r} \right) d\tau_{0},
\end{split}
\ee
where the integral is taken over a unit cell volume. The gradient in the Eq. (\ref{s_kp_2}) can be presented in the following way:
\be \label{kp_3}
\begin{split}
	K_{\nu \nu^{\prime}}^{+} &= \sum _{s, s^{\prime}} \int \left( \bm{e}_{ \bm{q} j } \cdot \bm{\nabla} V _{s^{\prime} s} \left(\bm{r}\right) \right) u_{ s \bm{k} }  ^ { \nu }  \left( \bm{r} \right) u_{ s^{\prime} \bm{k}^{\prime} } ^ { \nu^{\prime} \ast} \left( \bm{r} \right) d\tau_{0} 
	\\
	&=  \sum _{s, s^{\prime}}   \bm{e}_{ \bm{q} j } \cdot \int  \bm{\nabla}\left( u_{ s \bm{k} }  ^ { \nu } u_{ s^{\prime} \bm{k}^{\prime} } ^ { \nu^{\prime} \ast} V _{s^{\prime} s} \right) d \tau_{0} -   \sum _{s, s^{\prime}}  \bm{e}_{ \bm{q} j }  \cdot \int V_{s^{\prime} s} \bm{\nabla} \left( u_{ s \bm{k} }  ^ { \nu } u_{ s^{\prime} \bm{k}^{\prime} } ^ { \nu^{\prime} \ast} \right) d \tau_{0} ,
\end{split}
\ee
where $\bm{k} ^ {\prime} = \bm{k} + \bm{q}$. The first integral in (\ref{kp_3}) can be presented as a surface integral and vanishes because of the periodicity of the integrand. Then, 
\be \label{kp_3}
\begin{split}
	K_{\nu \nu^{\prime}}^{+} &= -   \sum _{s, s^{\prime}}  \bm{e}_{ \bm{q} j } \cdot \int V _{s^{\prime} s} \bm{\nabla} \left( u_{ s \bm{k} }  ^ { \nu } u_{ s^{\prime} \bm{k}^{\prime} } ^ { \nu^{\prime} \ast} \right) d \tau_{0}.
\end{split}
\ee
Using the $\bm{k}\cdot\bm{p}$ representation for $u_{ s \bm{k} }  ^ { \nu }$ and $u_{ s^{\prime} \bm{k}^{\prime} } ^ { \nu^{\prime} \ast}$ we arrive for the following equations for $u$-functions: 
\be \label{s_k_p_1}
	\begin{split}
		- \frac{\hbar ^ {2}}{2m} \nabla^2 u_{\uparrow \bm{k}} ^ {\nu}  + V_{\uparrow \uparrow} \left(\bm{r}\right) u_{\uparrow \bm{k}} ^ {\nu} + V_{\uparrow \downarrow} \left(\bm{r}\right) u_{\downarrow \bm{k}} ^ {\nu} - i \frac{\hbar ^ {2}}{m} \bm{k} \cdot \bm{\nabla} u_{\uparrow \bm{k}} ^ {\nu} = \left( \varepsilon ^ {\nu} \left( \bm{k} \right) - \frac{\hbar ^ {2} k ^ {2}}{2m}\right) u_{\uparrow \bm{k}} ^ {\nu}, 
		\\
		- \frac{\hbar ^ {2}}{2m} \nabla^2 u_{\downarrow \bm{k}} ^ {\nu}  + V_{\downarrow \uparrow} \left(\bm{r}\right) u_{\uparrow \bm{k}} ^ {\nu} + V_{\downarrow \downarrow} \left(\bm{r}\right) u_{\downarrow \bm{k}} ^ {\nu} - i \frac{\hbar ^ {2}}{m} \bm{k} \cdot \bm{\nabla} u_{\downarrow \bm{k}} ^ {\nu} = \left( \varepsilon ^ {\nu} \left( \bm{k} \right) - \frac{\hbar ^ {2} k ^ {2}}{2m}\right) u_{\downarrow \bm{k}} ^ {\nu}, 
		\\
		- \frac{\hbar ^ {2}}{2m} \nabla^2 u_{\uparrow \bm{k}^{\prime}} ^ {\nu^{\prime} \ast}  + V_{\uparrow \uparrow} ^{\ast} \left(\bm{r}\right) u_{\uparrow \bm{k}^{\prime}} ^ {\nu^{\prime} \ast} + V_{\uparrow \downarrow} ^{\ast} \left(\bm{r}\right) u_{\downarrow \bm{k}^{\prime}} ^ {\nu^{\prime} \ast} + i \frac{\hbar ^ {2}}{m} \bm{k} \cdot \bm{\nabla} u_{\uparrow \bm{k}^{\prime}} ^ {\nu^{\prime} \ast} = \left( \varepsilon ^ {\nu ^ {\prime} } \left( \bm{k} ^{\prime} \right) - \frac{\hbar ^ {2} k ^ {\prime 2}}{2m}\right) u_{\uparrow \bm{k}^{\prime}} ^ {\nu^{\prime} \ast}, 
		\\
		- \frac{\hbar ^ {2}}{2m} \nabla^2 u_{\downarrow \bm{k}^{\prime}} ^ {\nu^{\prime} \ast}  + V_{\downarrow \uparrow} ^{\ast} \left(\bm{r}\right) u_{\uparrow \bm{k}^{\prime}} ^ {\nu^{\prime} \ast} + V_{\downarrow \downarrow} ^{\ast} \left(\bm{r}\right) u_{\downarrow \bm{k}^{\prime}} ^ {\nu^{\prime} \ast} + i \frac{\hbar ^ {2}}{m} \bm{k} \cdot \bm{\nabla} u_{\downarrow \bm{k}^{\prime}} ^ {\nu^{\prime} \ast} = \left( \varepsilon ^ {\nu ^ {\prime} } \left( \bm{k}^{\prime} \right) - \frac{\hbar ^ {2} k ^ {\prime 2}}{2m}\right) u_{\downarrow \bm{k}^{\prime}} ^ {\nu^{\prime} \ast}.
	\end{split}
\ee
 $V_{\downarrow \uparrow} ^{\ast} \left(\bm{r}\right)  = V_{ \uparrow \downarrow } \left(\bm{r}\right) $ and $V_{ \uparrow \downarrow } ^{\ast} \left(\bm{r}\right)  = V_{\downarrow \uparrow} \left(\bm{r}\right)$ because $\hat{V}$ is a Hermitian operator. Multiplying the first equation in (\ref{s_k_p_1}) by $\bm{e}_{\bm{q}j} \cdot \bm{\nabla} u_{\uparrow \bm{k}^{\prime}} ^ {\nu^{\prime} \ast}$, the second one by $\bm{e}_{\bm{q}j} \cdot \bm{\nabla} u_{\downarrow \bm{k}^{\prime}} ^ {\nu^{\prime} \ast}$, the third one by $\bm{e}_{\bm{q}j} \cdot \bm{\nabla}u_{\uparrow \bm{k}} ^ {\nu} $, and the forth one by $\bm{e}_{\bm{q}j} \cdot \bm{\nabla} u_{\downarrow \bm{k}} ^ {\nu} $, adding them up with the successive integration over unit cell, we obtain:
\be \label{s_kp_3}
\begin{split}
&-  \bm{e}_{\bm{q}j} \cdot  \sum _{s, s^{\prime}}  \int V _{s^{\prime} s} \bm{\nabla}  \left( u_{ s \bm{k} }  ^ { \nu } u_{ s^{\prime} \bm{k}^{\prime} } ^ { \nu^{\prime} \ast} \right) d \tau_{0} = 
\\
&= \sum _{s} \int d \tau_{0} \left\{ - \frac{ \hbar^{ 2 } }{ 2 m }  \left[ \left(\bm{e}_{\bm{q}j} \cdot \bm{\nabla}  u_{s \bm{k}^{\prime}} ^ {\nu^{\prime} \ast}\right)  \nabla^{2}  u_{ s \bm{k}} ^ {\nu}  + \left(\bm{e}_{\bm{q}j} \cdot \bm{\nabla}  u_{s \bm{k}} ^ {\nu}\right) \nabla^{2}  u_{s \bm{k}^{\prime}} ^ {\nu^{\prime} \ast} \right]\right.  -
\\
&- i \frac{\hbar ^ {2} }{m} \left[ 
\left(\bm{e}_{\bm{q}j} \cdot \bm{\nabla}  u_{s \bm{k}^{\prime}} ^ {\nu^{\prime} \ast}\right) 
\left( \bm{k} \cdot \bm{\nabla}  u_{ s \bm{k}} ^ {\nu}\right) 
- 
\left(\bm{e}_{\bm{q}j} \cdot \bm{\nabla}  u_{s \bm{k}} ^ {\nu}\right) 
\left(\bm{k} \cdot \bm{\nabla}  u_{s \bm{k}^{\prime}} ^ {\nu^{\prime} \ast} \right)
\right] - 
\\
&
\left.
- \left[ \left( \varepsilon  ^ {\nu }  \left( \bm{k} \right) - \frac{\hbar ^ {2} k ^ {2}}{2m}\right) \left(\bm{e}_{\bm{q}j} \cdot \bm{\nabla}  u_{s \bm{k}^{\prime}} ^ {\nu^{\prime} \ast}\right) u_{ s \bm{k}} ^ {\nu}  + \left( \varepsilon^ {\nu ^ {\prime} } \left( \bm{k}^{\prime} \right) - \frac{\hbar ^ {2} k ^ {\prime 2}}{2m}\right) \left(\bm{e}_{\bm{q}j} \cdot \bm{\nabla}  u_{s \bm{k}} ^ {\nu}\right) u_{s \bm{k}^{\prime}} ^ {\nu^{\prime} \ast} \right]
\right\}
\end{split}
\ee

Due to periodicity of the integrands and using the Gauss' theorem, we transform Eq. (\ref{s_kp_3}) into a simpler form:
\be\label{s_kp_4}
\begin{split}
K_{\nu \nu^{\prime}}^{+} = & - \sum_{s}  i \frac{\hbar ^ {2} }{m}  \left(\bm{k} - \bm{k}^{\prime } \right) \cdot \int \left( \bm{\nabla}u_{ s \bm{k}} ^ {\nu} \right)  \left(\bm{e}_{\bm{q}j} \cdot \bm{\nabla}  u_{s \bm{k}^{\prime}} ^ {\nu^{\prime} \ast}\right) d\tau_{0} 
\\
&-  \sum_{s} \left[ \varepsilon  ^ {\nu }  \left( \bm{k} \right) - \varepsilon^ {\nu ^ {\prime} } \left( \bm{k}^{\prime} \right) - \left(  \frac{\hbar ^ {2} k ^ {2}}{2m} - \frac{\hbar ^ {2} k ^ {\prime 2}}{2m} \right)  \right]\int u_{ s \bm{k}} ^ {\nu}   \left(\bm{e}_{\bm{q}j} \cdot \bm{\nabla}  u_{s \bm{k}^{\prime}} ^ {\nu^{\prime} \ast}\right) d\tau_{0} .
\end{split}
\ee

We use the representation of $u_{ s \bm{k}} ^ {\nu}$ in terms of the Bloch periodic functions $\tilde{u}_{\bm{k}}$:\cite{anselm}
\be
\begin{split}
u_{ \uparrow \bm{k}} ^ {\nu}(\bm{r}) = a^{\nu}(k_z) \tilde{u}_{\bm{k} - \frac{\kappa}{2}\bm{e}_z}(\bm{r}) e^{-i\frac{\kappa}{2}z},\\
u_{ \downarrow \bm{k}} ^ {\nu}(\bm{r}) = b^{\nu}(k_z) \tilde{u}_{\bm{k} + \frac{\kappa}{2}\bm{e}_z}(\bm{r}) e^{+i\frac{\kappa}{2}z},
\end{split}
\ee
where $a^{\nu}$ and $b^{\nu}$ are found in Eq. (\ref{s_coef}). It is important to note that $\tilde{u}_{\bm{k}}$ is weakly dependent on a $\bm{k}$-vector. Indeed, in the free electron approximation employed in this work $\tilde{u} = 1 / \sqrt{ \Omega_0}$, where $\Omega_0$ is the unit cell volume. Then, we find the following expression for $K_{\nu \nu^{\prime}}^{+}$:
\be \label{s_kp_5}
\begin{split}
K_{\nu \nu^{\prime}}^{+} & = i \frac{\hbar ^ {2} }{m} \frac{1}{3} \left( \bm{q} \cdot \bm{e}_{\bm{q} j } \right) \left( a^{\nu} \left(\bm{k}\right) a^{\nu ^{\prime}} \left(\bm{k}^{\prime}\right) + b^{\nu} \left(\bm{k}\right) b^{\nu ^{\prime}} \left(\bm{k}^{\prime}\right)  \right) \int \left| \bm{\nabla} \tilde{u} \right|^{2} d\tau_{0}
\\
& + \frac{\hbar^{2} \kappa^{2}}{4m} q_{z} e_{\bm{q} j z} \left( a^{\nu} \left(\bm{k}\right) a^{\nu ^{\prime}} \left(\bm{k}^{\prime}\right) + b^{\nu} \left(\bm{k}\right) b^{\nu ^{\prime} } \left(\bm{k}^{\prime}\right)  \right) 
\\
& - i \frac{\kappa}{2} \left[ \varepsilon  ^ {\nu }  \left( \bm{k} \right) - \varepsilon^ {\nu ^ {\prime} } \left( \bm{k}^{\prime} \right) - \left(  \frac{\hbar ^ {2} k ^ {2}}{2m} - \frac{\hbar ^ {2} k ^ {\prime 2}}{2m} \right)  \right]  \left( a^{\nu} \left(\bm{k}\right) a^{\nu ^{\prime}} \left(\bm{k}^{\prime}\right) - b^{\nu} \left(\bm{k}\right) b^{\nu ^{\prime}} \left(\bm{k}^{\prime}\right)  \right) 
\end{split}
\ee

$K_{\nu \nu^{\prime}}^{-}$ corresponds  to $\bm{k}^{\prime}=\bm{k}-\bm{q}$ and differs only in the sign in $K_{\nu \nu^{\prime}}^{+}$.

For estimations $\int \left| \bm{\nabla} u \right|^{2} d\tau_{0} \approx 1/a^{2}$, where $a$ is the lattice constant. In the continuum model for the helicity potential $\kappa a / 2 \pi \ll 1$. Therefore, the second and the third terms in Eq. (\ref{s_kp_5}) can be neglected. Finishing up the electron part we obtain that:
\be \label{s_kp_6}
\begin{split}
K_{\nu \nu^{\prime}}^{+} & = i \frac{\hbar ^ {2} }{2 m a^{2} } \frac{2}{3} \left( \bm{q} \cdot \bm{e}_{\bm{q} j } \right) \left( a^{\nu} \left(\bm{k}\right) a^{\nu ^{\prime}} \left(\bm{k}^{\prime}\right) +b^{\nu} \left(\bm{k}\right) b^{\nu ^{\prime}} \left(\bm{k}^{\prime}\right)  \right).
\end{split}
\ee

Substituting Eq. (\ref{s_kp_6}) into the equation for the matrix element (\ref{s_mat_el_expanded}), we find:
\be \label{s_tme_1}
\matrixel{\bm{k}^{\prime}, \nu^{\prime}, N_{\bm{q}i}-1}{\Delta V}{\bm{k}, \nu, N_{\bm{q}i}} = - \frac{1}{\sqrt{N M}} K_{\nu \nu^{\prime}}^{+} \sqrt{\frac{ \hbar N_{\bm{q} j }}{ 2 \omega_{\bm{q} } }},
\ee
\be \label{s_tme_2}
\matrixel{\bm{k}^{\prime}, \nu^{\prime}, N_{\bm{q}i}+1}{\Delta V}{\bm{k}, \nu, N_{\bm{q}i}} = - \frac{1}{\sqrt{N M}} K_{\nu \nu^{\prime}}^{-} \sqrt{\frac{ \hbar \left( N_{\bm{q} j } + 1 \right) }{ 2 \omega_{\bm{q} } }},
\ee
where
\be \label{s_bose}
\left<N_{\bm{q}j}\right> = \frac{1}{e^{\frac{\varepsilon_{ph}}{k_{B}T}} - 1}
\ee
is the Bose-Einstein equilibrium distribution function.
As soon as transition matrix elements (\ref{s_tme_1}) (\ref{s_tme_2})  are determined, we substitute them into the scattering rates (\ref{s_w}) for the successive solution of the Boltzmann equation. It is important to note that $W_{\nu\nu^\prime}$ is a nondiagonal matrix allowing the transitions between the bands. 

 \be 
\begin{split}
W_{\bm{k} \bm{k}^{\prime}}^{\nu\nu^\prime}&=\frac{2\pi}{\hbar} \left|\matrixel{\bm{k}^{\prime}, \nu^{\prime}, N_{\bm{q}j}^{\prime}}{\Delta V}{\bm{k}, \nu, N_{\bm{q}j}}\right|^{2} \delta(\varepsilon_{i}(\bm{k}) - \varepsilon_{i^{\prime}}(\bm{k}^{\prime})) \\
&= \frac{2\pi}{\hbar} \frac{1}{N M} \frac{ \hbar N_{\bm{q} j }}{ 2 \omega_{\bm{q} } } \left| K_{\nu \nu^{\prime}}^{+} \right|^{2} \delta(\varepsilon_{\nu}(\bm{k}) - \varepsilon_{\nu^{\prime}}(\bm{k}^{\prime})) \delta(\bm{k}^{\prime} - \bm{k} - \bm{q}) \\
&+ \frac{2\pi}{\hbar} \frac{1}{N M} \frac{ \hbar \left( N_{-\bm{q} j } + 1 \right)}{ 2 \omega_{\bm{-q} } } \left| K_{\nu \nu^{\prime}}^{-} \right|^{2} \delta(\varepsilon_{\nu}(\bm{k}) - \varepsilon_{\nu^{\prime}}(\bm{k}^{\prime}))\delta(\bm{k}^{\prime} - \bm{k} + \bm{q}),
\end{split}
\ee
where $\bm{q} = \bm{k}^{\prime} - \bm{k}$ and $K_{\nu \nu^{\prime}}^{+}$ and $K_{\nu \nu^{\prime}}^{-}$ are determined by Eq. (\ref{s_kp_6}). These transition probabilities are used to solve the nonequilibrium Boltzmann equation. 

\bibliographystyle{iopart-num}
\section*{References}
\providecommand{\newblock}{}

\end{document}